**Journal of Geophysical Research: Space Physics**

**RESEARCH ARTICLE**
10.1002/2017JA024479# Hybrid Simulations of Positively and Negatively Charged Pickup Ions and Cyclotron Wave Generation at Europa

**Key Points:**
- A negatively charged pickup ion ring at Europa will generate near-circular right-hand polarized ion cyclotron waves consistent with theoretical predictions
- A trace negatively charged pickup ion population at Europa is able to generate observable right-handed wave power within a predominantly left-handed signal
- Hybrid simulations and linear theory support the hypothesis that the right-handed wave power observed by Galileo near Europa is due to the pickup of $Cl^-$

**Correspondence to:**
R. T. Desai,
r.t.desai@ucl.ac.uk

**Citation:**
Desai, R. T., Cowee, M. M., Wei, H., Fu, X., Gary, S. P., Volwerk, M., & Coates, A. J. (2017). Hybrid simulations of positively and negatively charged pickup ions and cyclotron wave generation at Europa. *Journal of Geophysical Research: Space Physics*, 122, 10,408–10,420. https://doi.org/10.1002/2017JA024479

Received 16 JUN 2017
Accepted 13 SEP 2017
Accepted article online 19 SEP 2017
Published online 26 OCT 2017R. T. Desai[1,2], M. M. Cowee[3], H. Wei[4], X. Fu[5], S. P. Gary[3,5], M. Volwerk[6], and A. J. Coates[1,2]

[1]Mullard Space Science Laboratory, University College London, London, UK, [2]Centre for Planetary Sciences, University College London/Birkbeck, London, UK, [3]Los Alamos National Laboratory, Los Alamos, NM, USA, [4]Institute of Geophysics and Planetary Physics, University of California, Los Angeles, CA, USA, [5]Space Science Institute, Boulder, CO, USA, [6]Space Research Institute, Austrian Academy of Sciences, Graz, Austria**Abstract** In the vicinity of Europa, Galileo observed bursty Alfvén-cyclotron wave power at the gyrofrequencies of a number of species including $K^+$, $O_2^+$, $Na^+$, and $Cl^+$, indicating the localized pickup of these species. Additional evidence for the presence of chlorine was the occurrence of both left-hand (LH) and right-hand (RH) polarized transverse wave power near the $Cl^+$ gyrofrequency, thought to be due to the pickup of both $Cl^+$ and the easily formed chlorine anion, $Cl^-$. To test this hypothesis, we use one-dimensional hybrid (kinetic ion, massless fluid electron) simulations for both positive and negative pickup ions and self-consistently reproduce the growth of both LH and RH Alfvén-cyclotron waves in agreement with linear theory. We show how the simultaneous generation of LH and RH waves can result in nongyrotropic ion distributions and increased wave amplitudes, and how even trace quantities of negative pickup ions are able to generate an observable RH signal. Through comparing simulated and observed wave amplitudes, we are able to place the first constraints on the densities of Chlorine pickup ions in localized regions at Europa.## 1. Introduction

The Alfvén-cyclotron instability is driven by a $T_\perp/T_\parallel > 0$ anisotropy in the distribution function of a given ion species, where $\perp$ and $\parallel$ are defined with respect to the ambient magnetic field. Anisotropic ion populations can be created by the ionization of neutral atoms where the newly formed ions are accelerated or "picked up" by an electric field. The ions are thus energized and, for a sufficiently high plasma beta, generate a number of electromagnetic microinstabilities and plasma wave phenomena (Gary & Schriver, 1987). Electromagnetic ion cyclotron waves (ICWs) associated with this ion pickup instability have been observed in the solar wind at comets, Mars, and Venus (Barabash et al., 1991; Delva et al., 2008; Thorne & Tsurutani, 1987), in the Earth's polar wind (Le et al., 2001), and within the Jovian and Kronian magnetospheres (Kivelson et al., 1996; Leisner et al., 2006). This article reports the first study addressing both positively and negatively charged pickup ions generating the Alfvén-cyclotron instability, with application to the Galilean moon Europa.

Within Jupiter's inner magnetosphere (<15 $R_J$), the magnetic field is approximately dipolar and nominally orientated at right angles to the sub-Alfvénic rotating magnetodisc. Ions picked up from satellites and rings in this environment will have low-velocity components parallel to the magnetic field and form perpendicular rings in velocity space unstable to the generation of left-hand (LH) polarized ICWs (Wu & Davidson, 1972). If, however, the newly ionized material has a significant velocity component parallel to the ambient magnetic field, the picked up ions will form ring-beam distributions in velocity space which results in a Doppler shift to the observed ICW resonant frequencies. In supra-Alfvénic plasma flows such as the solar wind, this can cause LH ICWs to be observed as right-hand (RH) polarized in the spacecraft frame (e.g., Jian et al., 2009; Wicks et al., 2016). It is also possible for obliquely propagating ICWs to become linearly polarized and to then reverse polarization if they attain and pass through the crossover frequency of the multicomponent plasma in which they are generated (Petkaki & Dougherty, 2001; Rauch & Roux, 1982).

©2017. The Authors.
This is an open access article under the terms of the Creative Commons Attribution License, which permits use, distribution and reproduction in any medium, provided the original work is properly cited.DESAI ET AL.    ICWS AT EUROPA    10,408

These gyroresonant transverse electromagnetic fluctuations have been observed to scatter pickup ion distributions into a bispherical shell distribution and toward thermal equilibrium (Coates et al., 1990; Tokar et al., 2008), with energy distributed between wave growth and ion heating (Cowee et al., 2007; Huddleston et al., 1998).

The moon Europa orbits Jupiter at ∼8.8 $R_J$ and has been identified as the secondary mass loading source in the Jovian magnetosphere, contributing an estimated 1–100 kg/s of neutral material (Kivelson et al., 2009; Shemansky et al., 2014). Galileo observed a significant amount of plasma wave activity in the vicinity of the moon (Kurth et al., 2001; Volwerk et al., 2001) and, during one upstream encounter and two through the moon's plasma wake, observed bursty ICW characteristics at the gyrofrequencies of a number of species including $K^+$, $Na^+$, $O_2^+$, and $Cl^+$. A notable trend within this data set was the occurrence of both LH and RH polarized wave power at the $Cl^+$ gyrofrequency, a phenomenon absent at the gyrofrequency of other minority species picked up locally at the moon. These waves were highly field aligned, and the RH wave power was consequently hypothesized to result from the pickup of both positively and negatively charged chlorine ions, possible due to the high electron affinity (3.61 eV) and stable configuration of the chlorine anion, $Cl^-$ (Kivelson et al., 2009; Volwerk et al., 2001).

Negatively charged ions have been observed being picked up from the icy moon Rhea (Teolis et al., 2010), from Saturn's main rings (Jones & Coates, 2014), in the Enceladus plumes (Coates et al., 2010), and in Titan's ionosphere (Coates et al., 2007; Desai et al., 2017) where the outflow of these species is also predicted (Ledvina & Brecht, 2012). Negatively charged ions have also been observed in abundance at 1P/Halley and 67P/Churymov-Gerasimenko (Burch et al., 2015; Chaizy et al., 1991) and likely form significant plasma populations in the outer solar system where the plasma temperatures are typically lower (lower associative detachment rates) and the solar photon flux is considerably reduced (lower photodetachment rates). The physics of ICWs generated by negatively charged pickup ions has, however, not been examined. The presence of chlorine pickup ions at Europa also suggests the moon to be a net source of this species with implications for the abundance of NaCl in the subsurface ocean.

This work examines the physics of ICWs generated by both positively and negatively charged pickup ions with application to the Europan plasma environment. Section 2 describes a self-consistent hybrid simulation approach which is used to study the growth and nonlinear evolution of the modes generated in the plasmas of interest. The analysis first looks to characterize the behavior of the negative ion ring instability using linear dispersion theory in section 3 and hybrid simulations in section 4.1, which are contrasted to the well-known characteristics and properties of the positive ion ring instability. Sections 4.2 and 4.3 go on to study scenarios of both instabilites generated within the same system to characterize any interaction between the two and their observable signatures. Section 5 then describes a parametric study of the instability saturation energy with ring properties such as density and anisotropy with close reference to the Galileo magnetometer observations.

## 2. Methodology

Warm plasma dispersion theory and hybrid simulations have previously constrained the behavior of the Alfvén-cyclotron instability generated by heavy $SO_2^+$ ion rings in the Io plasma torus (Cowee et al., 2006; Huddleston et al., 1997, 1998) and also for water group pickup ion rings within Saturn's extended neutral cloud (Cowee et al., 2009; Leisner et al., 2006; Rodríguez-Martínez et al., 2010). This study focuses on the Europan plasma environment and draws comparisons to these studies.

The hybrid code has successfully reproduced the Alfvén-cyclotron instability for a number of plasma environments (e.g., Cowee et al., 2006; Gary et al., 1989; Omidi et al., 2010). The code specifies ions kinetically to capture phenomena occurring at ion spatial and temporal scales and approximates electrons as a massless neutralizing fluid (Winske et al., 1992, 2003). The simulations consist of one spatial dimension, $x$, aligned with the ambient magnetic field, $\mathbf{B}_0$, a setup justified as the main wave power is expected at parallel propagation (Wu & Davidson, 1972). All three components of electromagnetic fields and velocities are resolved to capture the predominantly transverse electromagnetic fluctuations along the $y$ and $z$ axes. The simulations are carried out in the plasma frame with zero-drift velocity, and the particle-in-cell simulation technique is employed where all ions are represented by superparticles whose densities and currents are collected on a spatial grid and used as sources to the field equations. Maxwell's equation are implemented in the Darwin limit where the displacement current, $\partial \mathbf{E}/\partial t$, is neglected within Ampére's law to eliminate high-frequency perturbations

**Table 1**
*Nominal Plasma Parameters for All Simulations Runs Unless Otherwise Stated*

| $j$ | $m_j/m_p$ | $q_j/q_p$ | $T_\parallel$ (eV) | $T_\perp$ (eV) | $v_{ring}$ (km s$^{-1}$) |
|---|---|---|---|---|---|
| Light ion core | 8 | +1 | 100 | 100 | - |
| Cl$^-$ core | 35 | −1 | 100 | 100 | - |
| Cl$^+$ core | 35 | +1 | 100 | 100 | - |
| Cl$^-$ ring | 35 | −1 | ∼0 | 1830 | 100 |
| Cl$^+$ ring | 35 | +1 | ∼0 | 100 | 100 |

*Note.* $B_0$ = 400 nT; $n_0$ = 100/cm$^3$; $c/\omega_{pi}$ = 134.8 km; $\Omega_i/2\pi$ = 0.17 Hz; $c/v_A$ = 344.

such as light waves (Darwin, 1920). In these initial value simulations the number of superparticles are specified at the beginning of the run and held as constant, in order to represent an injection of pickup ions from Europa, all particles are initialized with a quiet start configuration to minimize any net currents at the start of the run.

The inputs into the simulation are representative of the general plasma environments around Europa with emphasis on the eleventh (E11) and fifteenth (E15) Galileo encounters. A variety of pickup velocities, from 55 to 100 km/s, are explored to examine the scaling of the instability with respect to inherent variations of this parameter due to effects such as the slowdown of the incident magnetoplasma in front of the moon, the formation of Alfvén wings, pickup ion currents, wake effects, and the flapping of the Jovian magnetodisk (Khurana et al., 1998; Neubauer, 1998, 1999). Galileo observed electron densities from 30/cm$^3$ up to several 100/cm$^3$ during the various encounters, an average ion charge of 1.5 $q_e$ and a mean ion mass of 18.5 unified atomic mass units (u) dominated by low-charge states of iogenic sulfur and oxygen (Bagenal et al., 2015; Kivelson et al., 2009). A nominal ion density of 100/cm$^3$ and magnetic field of $\mathbf{B}_0$ = 400 nT are taken as representative of the Galileo observations during E11 and E15. Further, simulation parameters are outlined in Table 1 which are used in all runs unless otherwise stated.

The plasma is considered to consist of multiple ion species, either in a zero-drift Maxwellian velocity distribution or a delta function velocity ring distribution with $T_\parallel \sim 0$. The $\delta\mathbf{E}$ perturbations of the transverse waves act to scatter the ring ions in directions perpendicular to $B_0$, while the $\mathbf{v} \times \delta\mathbf{B}$ forces scatters them in the directions parallel and antiparallel to $B_0$. Any other ion population with a similar mass per charge ratio will gyroresonantly interact with these waves and be subject to the same forces and must also be included in the simulations. Ion species with a different mass per charge ratio are not expected to interact with the instability and are represented by a light ion core population to maintain charge neutrality. Each species has mass, $m_j$, and charge state, $q_j$, normalized to the proton scale with parallel and perpendicular velocities, $V_{\parallel j} = (2k_B T_{\parallel j}/m_j)^{1/2}$ and $V_{\perp j} = (2k_B T_{\perp j}/m_j)^{1/2}$. $V_\perp$ is set as the pickup velocity, which is defined as the difference between the near-corotational background plasma flow at the orbit of Europa of ≤117 km/s and the moons orbital velocity of ∼14 km/s. $V_\parallel$ is set to nearly zero, under the assumption of minimal relative motion between the atmospheric neutrals and the moon.

The temperature anisotropy is defined as $A_j = T_{\perp j}/T_{\parallel j}$, the Alfvén velocity as $v_A = B_0/\sqrt{\mu_0 n_0 m_j}$, the ion gyrofrequency as $\Omega_j = q_j B_0/m_j$, the plasma frequency as $w_{pi}^2 = n_0 q_i^2/\epsilon_0 m_j$, and the ion parallel plasma beta as $\beta_{\parallel j} = 2\mu_0 n_0 T_{\parallel j}/B_0^2$ where the fluid electron beta is equal to the sum of the ion components, $\beta_e = \sum \beta_j$. The subscript $p$ denotes protons, $e$ denotes electrons, and $i$ denotes chlorine ions. The simulation outputs are given in normalized units of the ion Cl$^+$/Cl$^-$ gyrofrequency $\Omega_i$ where $\Omega_i/2\pi$ = 0.17 Hz, the ion inertial length where $c/\omega_{pi}$ = 134.8 km, and the Alfvén velocity where $v_A$ = 147.5 km/s. The simulation domain is 512 $c/\omega_{pi}$ in length with 512 grid cells with periodic boundary conditions, and between 50 and 100 superparticles per cell are used within the various runs. The runs are then stepped forward in time steps of 0.05 $\omega_{pi}^{-1}$, until some time after the instability has reached a quasi-steady saturation energy level.

## 3. Linear Dispersion Theory

Figure 1 illustrates solutions to the kinetic linear dispersion equation for electromagnetic fluctuations at $\mathbf{k} \times \mathbf{B}_o$ = 0 in a homogeneous, magnetized, collisionless plasma with three components: electrons and protons, each with Maxwellian velocity distributions, and a singly charged helium ion or anion component with a cold velocity ring distribution (e.g., see Gary & Madland, 1988, with $\alpha = 90°$). The pickup or ring velocity to Alfvén velocity ratio is specified as $\mathbf{v_r}/\mathbf{v_A} = 0.5$ and the parallel beta of the proton is $\beta_\parallel = 0.15$. These are in the range of observed values, with variations around these parameters not significantly altering the calculated solutions. Although the simulations described in subsequent sections address much heavier ions, we expect that the qualitative features of the various modes shown here should be similar to those of the simulations.

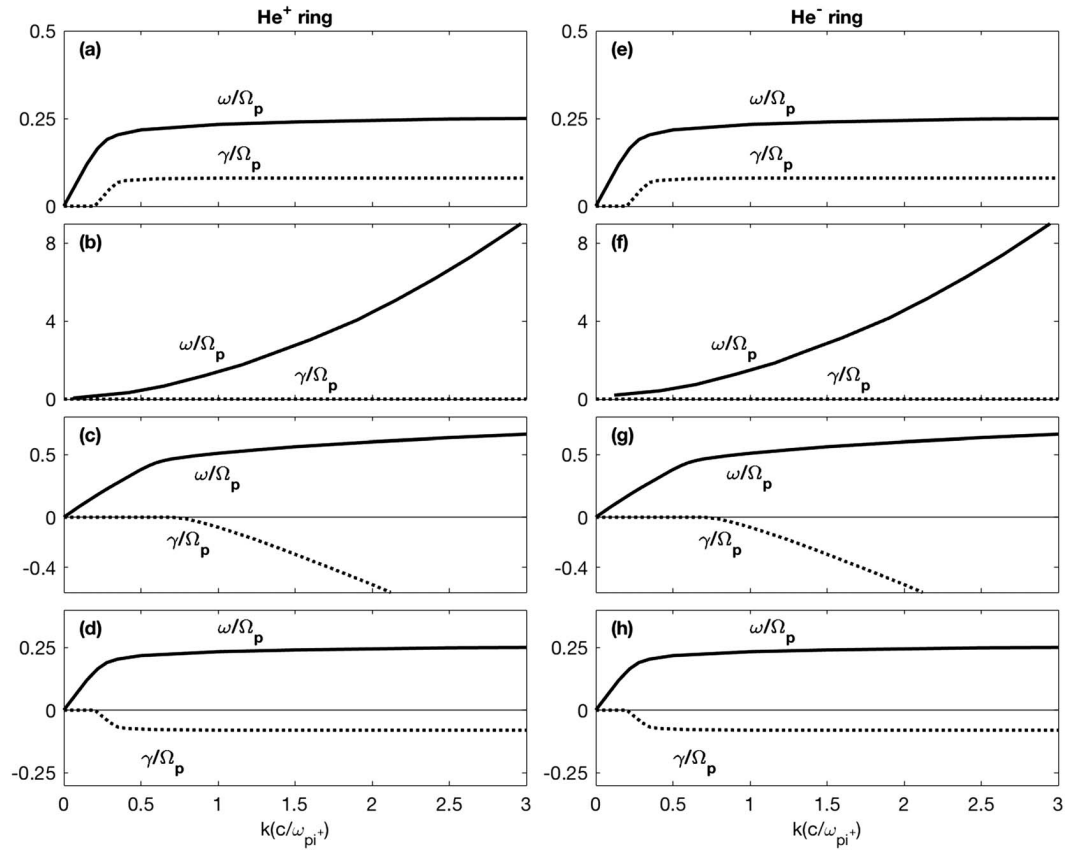

**Figure 1.** Solutions to the kinetic linear dispersion equation for electromagnetic fluctuations at $\mathbf{k} \times \mathbf{B}_o = 0$ in a homogeneous, magnetized, collisionless plasma. There are three components to the plasma: electrons and protons, each with Maxwellian velocity distributions, and a singly charged helium ion or anion component with a cold velocity ring distribution. Here $n_p/n_e = 0.95$ and $n_{He}/n_e = 0.05$. The ring velocity is $v_{ring}/v_A = 0.50$ and the ion beta value (based on the electron density and the proton temperature) is 0.15. Here the solutions are normalized to the proton gyrofrequency, $\Omega_p$. (a–d) The results for He$^+$ ions; (e–h) the results for He$^-$ anions. Figures 1a and 1e show dispersion of the left-hand and right-hand helium cyclotron instabilities, respectively; Figures 1b and 1f illustrate dispersion of the stable right-hand polarized magnetosonic modes, Figures 1c and 1g represent the left-hand polarized stable Alfvén-cyclotron modes, and Figures 1d and 1h illustrate the left-hand and right-hand polarized stable helium cyclotron modes, respectively.

Plasma parameters used in Figure 1 are given in the figure caption. The complex frequency is $\omega = \omega_r + i\gamma$ with positive $\gamma$ corresponding to fluctuation growth. The left-hand column shows results for a He$^+$ ion velocity ring, whereas the right-hand column illustrates results for a He$^-$ anion velocity ring. In both cases the dispersion equation yields four distinct modes near or below the cyclotron frequency with both positive and negative helicities. Magnetic helicity defines the sense of the spatial rotation of the fluctuating field vectors with respect to the wave vector, $\mathbf{k}$, at a given instance and is independent of the reference frame of the observer (Gary, 1993). A positive helicity corresponds to either a forward propagating RH wave or a backward propagating LH wave and a negative helicity corresponds to either a backward propagating RH or forward propagating LH wave.

The fastest growing modes in both cases are the helium cyclotron instabilities shown in Figures 1a and 1e. The figure suggests that these two modes are unstable to arbitrarily short wavelengths, however, for the more realistic case of a warm ring ($T_\parallel > 0$) ion cyclotron resonances will damp modes at short wavelengths (e.g., Figure 2 of Gary & Madland, 1988) so that we expect maximum growth to arise near $kc/\omega_{pi}$. In addition

**Table 2**
*Case Study Simulations for Runs I–IV Described in the Article Text*

| Run | Components | $n_i$ (1/cm$^3$) | | |
|---|---|---|---|---|
| I | Light ion core | 80 | | |
|  | Cl$^-$ core | 15 | | |
|  | Cl$^-$ ring | 5 | | |
| II | Light ion core | 60 | | |
|  | Cl$^+$ core | 15 | | |
|  | Cl$^-$ core | 15 | | |
|  | Cl$^+$ ring | 5 | | |
|  | Cl$^-$ ring | 5 | | |
|  |  | (Both) | (Cl$^+$) | (Cl$^-$) |
| III | Light ion core | 98 | 98.5 | 99.5 |
|  | Cl$^+$ ring | 1.5 | 1.5 | - |
|  | Cl$^-$ ring | 0.5 | - | 0.5 |
| IV | Light ion core | 98 | 98.1 | 99.9 |
|  | Cl$^+$ ring | 1.9 | 1.9 | - |
|  | Cl$^-$ ring | 0.1 | - | 0.1 |

*Note.* Further plasma input parameters are given in Table 1.

to these, Figures 1b and 1f show the right-handed magnetosonic whistler mode that is undamped, and Figures 1c and 1g show the Alfvén-cyclotron modes which undergo strong proton cyclotron damping at short wavelengths, and Figures 1d and 1h which illustrate the helium cyclotron waves which also become damped at sufficiently short wavelengths.

## 4. Results From Hybrid Simulations
### 4.1. Negative Ring Instability

In simulation Run I, the instability is simulated with negatively charged Cl ions. As outlined in Table 2, a cold Cl$^-$ ring distribution is initialized with negligible temperature spread parallel to the magnetic field together with a thermalized Cl$^-$ core population. A further positively charged Maxwellian light ion core represents the ambient low-charge state oxygen and sulfur background ions of the Jovian magnetodisk. This background population was initially specified as protons although a higher mass ion was later used so that larger time steps could be taken while not affecting the growth of the instability and still accurately resolving the ion gyromotion. Figure 2 shows the fluctuating magnetic field energy density, $(\delta B/B_0)^2$, and temperature evolution of the different species. The instability initially grows rapidly until $\Omega_i t = 60$–70 then grows slowly until $\Omega_i t = 110$–120 where a quasi-steady level of $(\delta B/B_0)^2 = \sim 2.7 \times 10^{-4}$ is reached.

The fluctuations shown in Figure 2a act to pitch angle scatter the anisotropic Cl$^-$ ring distribution toward isotropy as shown in Figures 2c–2e. The Cl$^-$ core population acts to damp the growth of the waves and is consequently heated in the perpendicular direction, while the light ion core remains at a constant temperature as anticipated for components not resonantly interacting with the instability.

Figure 3 shows the $\omega$-$k$ spectrum during the growth phase of the instability. The wave power grows out of the Alfvén branch and is concentrated near the chlorine cyclotron frequency with the bulk wave power occurring between $kc/\omega_{pi} = 1$–2. This agrees with the dispersion relations given in Figure 1 where positive growth rates and similar frequencies are predicted near $kc/\omega_{pi}$. The dynamic spectrum in Figure 4 shows the power spectral density (PSD) of Run I and the magnetic ellipticity as calculated by the UCLA X waves analysis tool, which utilizes the inherently antisymmetric quadrature spectral matrix to provide information on the propagation direction and thus polarization of separated circularly polarized wave components (Means, 1972). In the case of superimposed waves with different polarizations, this technique will return the net resultant ellipticity. This was generated using a time series determined from one grid cell in the simulation and while other grid cells may yield slightly different results, the simulation conditions are initially uniform and so significant differences between cells are not expected. The wave power occurs at the Chlorine gyrofrequency as in Figure 3 but decreases slightly in frequency over time as a result of an inverse cascade to longer wavelengths (Gary & Madland, 1988). The variability in the wave power is due to the superposition of a range of modes excited in the simulation. The resultant ellipticities appear close to +1, indicating a RH near circularly polarized wave near the chlorine gyrofrequency as predicted by the positive helicity mode in Figure 1.

The results of the negative ion ring instability presented in Figures 2–4 show that a negatively charged ring generates a RH instability in-line with the predictions of linear theory and is analogous to that of the LH instability studied in Huddleston et al. (1997) and Cowee et al. (2006).

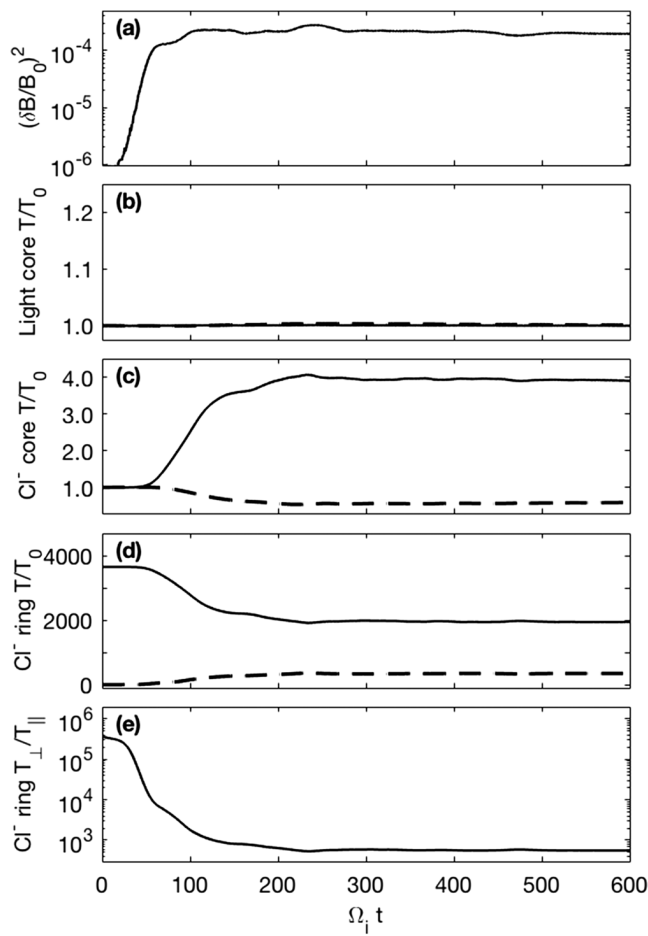

**Figure 2.** Time histories from Run I which includes light ion core, Cl$^-$ core, and Cl$^-$ ring ion components. (a) The fluctuating magnetic field energy density, (b–d) the perpendicular (solid line) and parallel (dashed line) temperature histories of the different components, and (e) the anisotropy of the ring ions. Input parameters are provided in Tables 1 and 2.

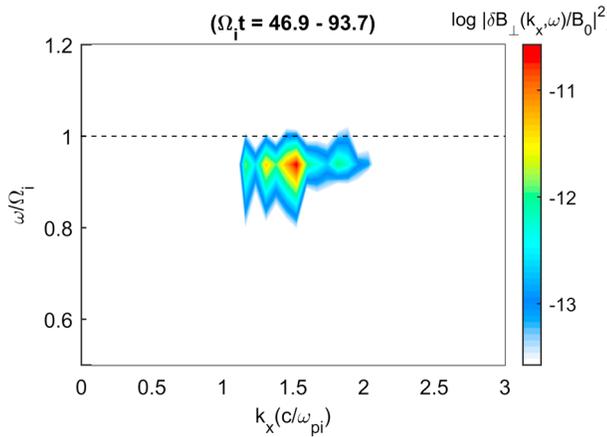

**Figure 3.** Frequency-wave number spectrum of the transverse wave power from Run I which includes light ion core, Cl$^-$ core, and Cl$^-$ ring ion components. Waves propagate in both directions with the same spectral properties. The wave power is seen near the Cl$^-$ gyrofrequency (dashed line) and at normalized wave numbers between 1 and 2. The spectrum is calculated using a Fourier window over $46.9 < \Omega_i t < 93.7$.

### 4.2. Simultaneous Pickup of Positive and Negative Ions

This section examines the cogeneration and interaction of both positive (LH) and negative (RH) Alfvén-cyclotron instabilities within the same system. To explore these effects, Run II contains equal quantities of Cl$^+$ and Cl$^-$ ring and core ions as outlined in Table 2. This results in twice the amount of free energy available to drive wave growth, and, in the absence of interactions between the two rings, the saturation energy should reach double that of Run I (Figure 2a) where only the negative ring ions are initialized.

Figure 5 shows the fluctuating magnetic field energy density growing as anticipated although in this instance intense high-frequency perturbations are apparent. These oscillations cause the magnetic field energy density, $(\delta B/B_0)^2$, to reach levels of up to 4 times that of Run I, although a running average produces the anticipated saturation energy level of $(\delta B/B_0)^2 = \sim 5.4 \times 10^{-4}$, twice that of Run I. These high-frequency oscillations occur at a frequency of approximately twice the chlorine ion gyrofrequency although the intensity also changes in a further beating pattern along a time scale of $\Omega_i t = 25-50$ indicating the superposition of multiple modes. The high-frequency oscillations are also apparent in the temperature histories of the Chlorine core components and to a lesser extent in the nonresonant light ion core. To further examine this behavior, the phase space angular distributions of the different components are plotted in Figure 6. Here both the Cl$^+$ and Cl$^-$ core and ring components can be seen to be bunched in gyrophase space with particles clustered around common positions while gyrating about $B_0$.

Nongyrotropic ion distributions have been observed at Comet 26P/Grigg-Skjellerup (Coates et al., 1993) and in the distant Earth's magnetotail (Saito et al., 1994). Subsequent studies on these determined that increasing the nongyrotropy of anisotropic ion distributions resulted in increased instability growth rates and a larger range of excited wave numbers (Brinca et al., 1993; Convery et al., 2002). These studies also show, using hybrid simulations, how nongyrotropic ring distributions result in large magnetic oscillations as energy is exchanged between the waves and the time-varying bunched ion distributions. These effects appear analogous to what is seen in Run II although in this study the ions are initialized in a gyrotropic state and develop their nongyrotropy as the simulation progresses. Density oscillations at half the wavelength of the ICW mode have previously been observed in simulations of anisotropic ring current ions in the Earth's magnetosphere (Omidi et al., 2010). These were attributed to develop from oppositely directed ICWs intersecting one another twice within each full wave rotation and introducing periodic effects at twice the resonant wave frequency. This caused the core ions to become bunched in gyrospace as seen in Run II although an associated electrostatic periodicity was also reported which is not observed in this study. The half-length oscillatory behavior seen in Run II also appears significantly greater than that reported in Omidi et al. (2010), presumably due to the presence of both LH and RH waves acting to bunch both positively and negatively charged core components. The cogeneration of the LH and RH instabilities could also introduce multiple half-length modes which will naturally vary over time moving in and out of phase and could explain the large variations in intensity.

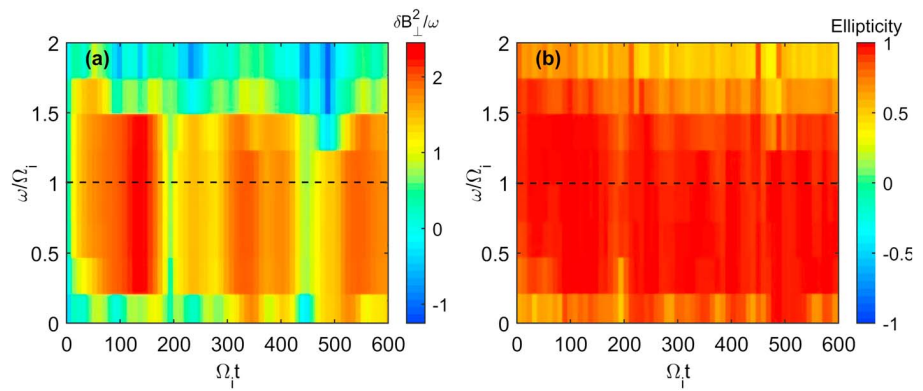

**Figure 4.** Run I dynamic spectrum of (a) transverse power spectral density near the Cl$^-$ gyrofrequency (dashed line) and (b) magnetic polarization. The corresponding magnetic polarization in Figure 4b is close to +1, indicating RH polarized wave power.

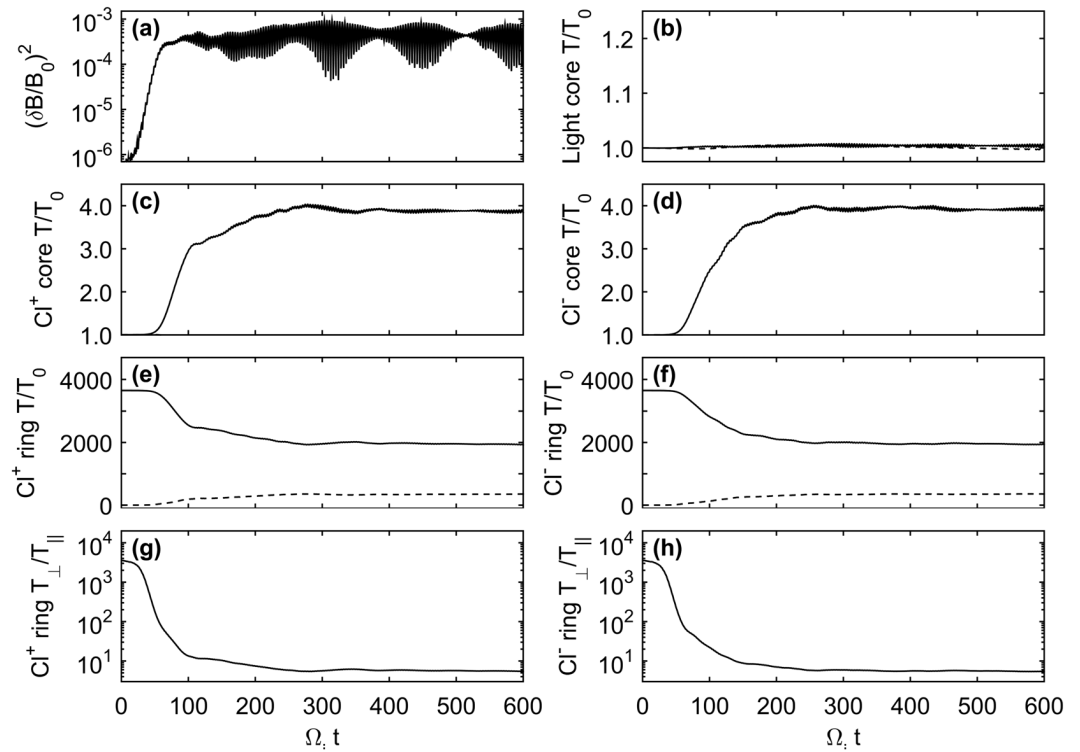

**Figure 5.** Time histories from Run II which includes light ion core, Cl$^-$, and Cl$^+$ ion cores as well as Cl$^+$ and Cl$^-$ ring ion components. (a) The fluctuating magnetic field energy density, (b–f) the perpendicular (solid line) and parallel (dashed line) temperature history of the different components, and (g–h) the anisotropy of the positive and negative ring ions, respectively. Input parameters are given in Tables 1 and 2.

Figure 7 displays the $\omega$-$k$ spectrum for Run II and shows that the wave power is spread over wave numbers and frequencies similar to those of Run 1 (Figure 2a), although in this instance there is increased wave power due to double the density of ring ions included. The dynamic spectra in Figure 8 also shows the wave power appearing at and just below the chlorine ion gyrofrequency. The wave power, however, displays no clear circular polarization with ellipticity values sometimes displaying positive or negative values. This effective near-linear polarization agrees with the theoretical expectation of the summation of LH and RH circularly polarized waves at similar frequencies and wave numbers. The variability in the polarization is attributed to the time-dependent fluctuating wave amplitudes, frequencies, and wave numbers for each instability which allows either the LH or the RH mode to appear dominant for short intervals. This effect could also be amplified as a result of the gyrophase bunching.

### 4.3. The Magnetic Signature of Negative Ion Pickup

This section describes an ensemble of simulations which further characterize the interaction and scaling of the two instabilities and explore how a trace negative ion population is able to generate an observable signal. These runs cover a wide range of Cl$^+$ and Cl$^-$ densities, and two specific runs are chosen, termed Run III and Run IV in this article, to illustrate the key findings. Run III and Run IV nominally contain a 3:1 and 19:1 Cl$^+$:Cl$^-$ pickup ion density ratio, respectively, as outlined in Table 2. Figure 9 shows the fluctuating magnetic field energy density, $\omega$-$k$ spectrum, and magnetic polarization for both Run III and Run IV. The Cl$^+$ and Cl$^-$ instabilities are also simulated individually and the results overlaid.

In Runs III and IV the Cl$^+$ instability has an increased growth rate and wave amplitudes compared with the significantly reduced Cl$^-$ instability and the sum of the LH and RH instabilities are in good agreement with the amplitudes produced when both components are included. In $\omega$-$k$ space the LH and RH instabilites appear as distinct pockets of wave power at different wave numbers. The wave power still occurs near the chlorine ion gyrofrequency, but the Cl$^-$ instabilities are excited at slightly different frequencies to the Cl$^+$ instabilities in both instances.

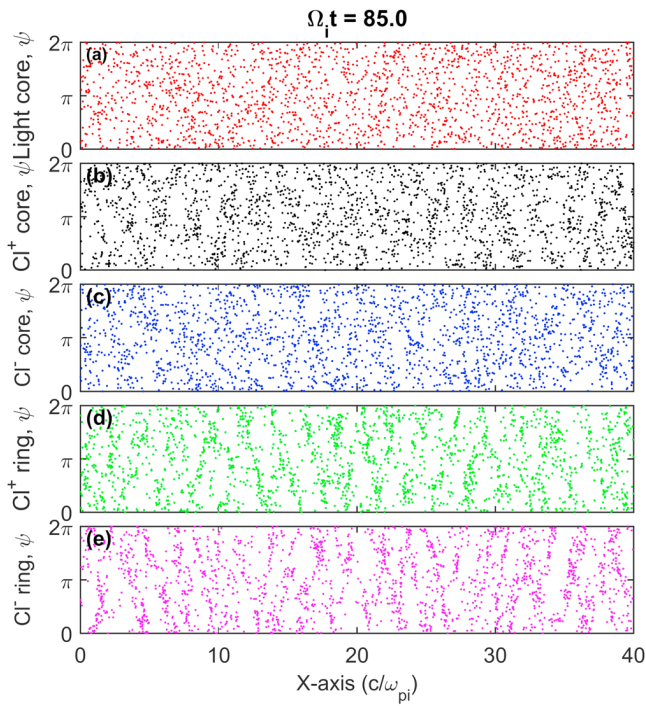

**Figure 6.** Run II plasma component distributions relative to their gyrophase angle, $\psi$, at $\Omega_i t = 85.0$. Bunching is present in all four chlorine ion components due to counterpropagating ICWs.

During Run III the magnetic polarization near the chlorine ion gyrofrequency is predominantly LH polarized although, at $\Omega_i t = 200$, when the weaker Cl$^-$ RH instability amplitudes become comparable and briefly exceed those of the Cl$^+$ LH instability, a polarization reversal occurs and the ICW power becomes predominantly RH polarized before returning to a LH polarization. In Run IV, however, despite the Cl$^-$ wave amplitudes never exceeding those generated by the Cl$^+$ ring, a polarization reversal is surprisingly also present and a brief burst of RH polarized wave power results at $\Omega_i t = 250$. This phenomenon is attributed to the two instabilities having different growth rates and generating wave power at different wavelengths and frequencies. The $\omega$-$k$ spectrum indicates that during this brief instance of a RH-dominated spectrum, the RH wave power is generated very close to the chlorine ion gyrofrequency, whereas the LH instability saturates earlier and although possessing more wave power, this wave power is spread across a wider range of wave numbers and centers at a lower frequency. This effect means that despite the reduced growth rate of the Cl$^-$ instability, the wave power is still able to generate an observable RH signal as a result of the natural time-dependent variability of the wave amplitudes.

The high-frequency oscillatory behavior is also reduced in Run III and even more so in Run IV, presumably due to the LH and RH components occurring at different wave numbers. This suggests that this half-length mode is more strongly driven when LH and RH ICWs of a similar wave number and frequency are present.

The results obtained from Run III and Run IV demonstrate how positive and negative ring instabilities, although similar in nature, can have different spectral properties as a result of differences in the initial ion distribution functions. The simulations thus show how the appearance of RH wave power within a predominantly LH signal, as was observed by Galileo, can be evidence of negatively charged pickup ions generating a RH Alfvén-cyclotron instability. The simulations also indicate how an inherent polarization variability will be present in systems containing instabilities driven by both positive and negative ion rings.

## 5. Application to Europa

To better understand the processes occurring at Europa, the simulated waves are compared to the ICWs observed by Galileo. The dominant pickup ion at Europa is $O_2^+$ produced from the moon's oxygen-based exosphere (Hall et al., 1995; Johnson et al., 2009), but large concentrations of thermalized $S^+$ ions originating from Io have a similar mass per charge ratio and resonantly interact to damp the growth of waves near the $O_2^+$ gyrofrequency. Consequently, $O_2^+$ ICWs are not observed in significant quantities and during the E11 encounter these waves appear to be generated in a higher-frequency band to those concentrated near the Cl$^{+/-}$ gyrofrequency. During E15, however, it is difficult to rule out LH wave power from ICWs generated by the pickup of $O_2^+$ (see Volwerk et al., 2001, Plates 1 and 2).

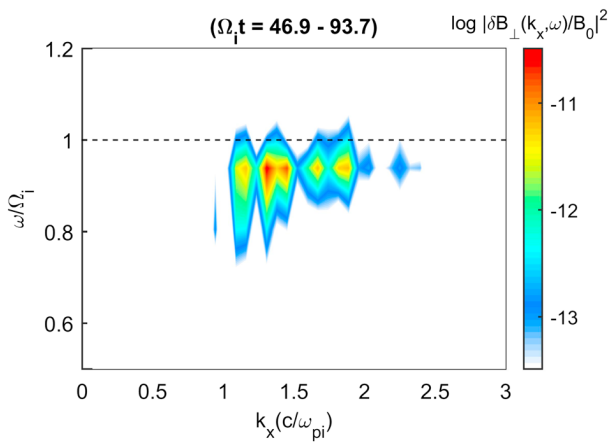

**Figure 7.** Run II frequency-wave number spectrum of the transverse wave power where equal quantities of positive and negative ring ions are included. Waves propagate in both directions with the same spectral properties near the Cl$^-$ gyrofrequency (dashed line) at normalized wave numbers between 1 and 2. The spectrum is calculated using a Fourier window over $46.9 < \Omega_i t < 93.7$.

The E11 and E15 observations of wave power at the chlorine gyrofrequency occurred between 3 and 4 Europa radii ($R_{Eur} \sim 1561$ km) downstream of the moon (Volwerk et al., 2001). The simulation results presented in Runs I–IV show, for the various densities examined, the instability saturates within 200–320 $\Omega_i^{-1}$, which, in the plasma frame, corresponds to a convection distance of $<1$ $R_{Eur}$. The simulations therefore suggest that the Cl$^+$ and Cl$^-$ ring instabilities would likely have reached saturation when observed by Galileo, if the ions were picked up close to the moon.

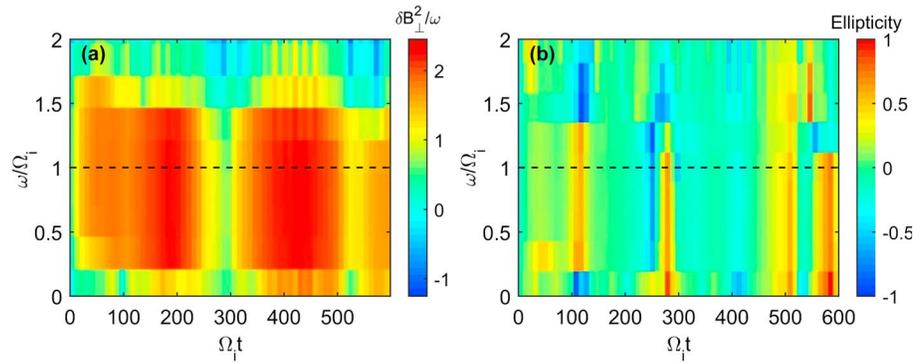

**Figure 8.** Run II dynamic spectrum for both the Cl$^+$ and Cl$^-$ instabilities. (a) The transverse power spectral density and (b) magnetic polarization, near the chlorine ion gyrofrequency (dashed black line). The wave power is observed in the same regions as in Figure 3, and the magnetic polarization is approximately linear as anticipated from the summation of the LH and RH near circularly polarized components although periods of both LH and RH polarizations are present.

To relate possible source densities to the Galileo observations, a parametric study of the Alfvén-cyclotron instability is carried out for magnetic field fluctuations in the range observed by Galileo; see Table 3. During the E11 and E15 encounters the wave amplitudes at the Cl$^{+/-}$ gyrofrequency reached absolute values of 2.5 nT, $(\delta B/B_0)^2 = 3.9 \times 10^{-5}$, and 3.5 nT, $(\delta B/B_0)^2 = 7.7 \times 10^{-5}$, respectively. The wave saturation amplitude is necessarily dependent on the background density of Cl$^{+/-}$ ions which will act to damp the growth of the instability. In Figure 10 the ring ion densities are plotted against saturated amplitude for the two limiting cases

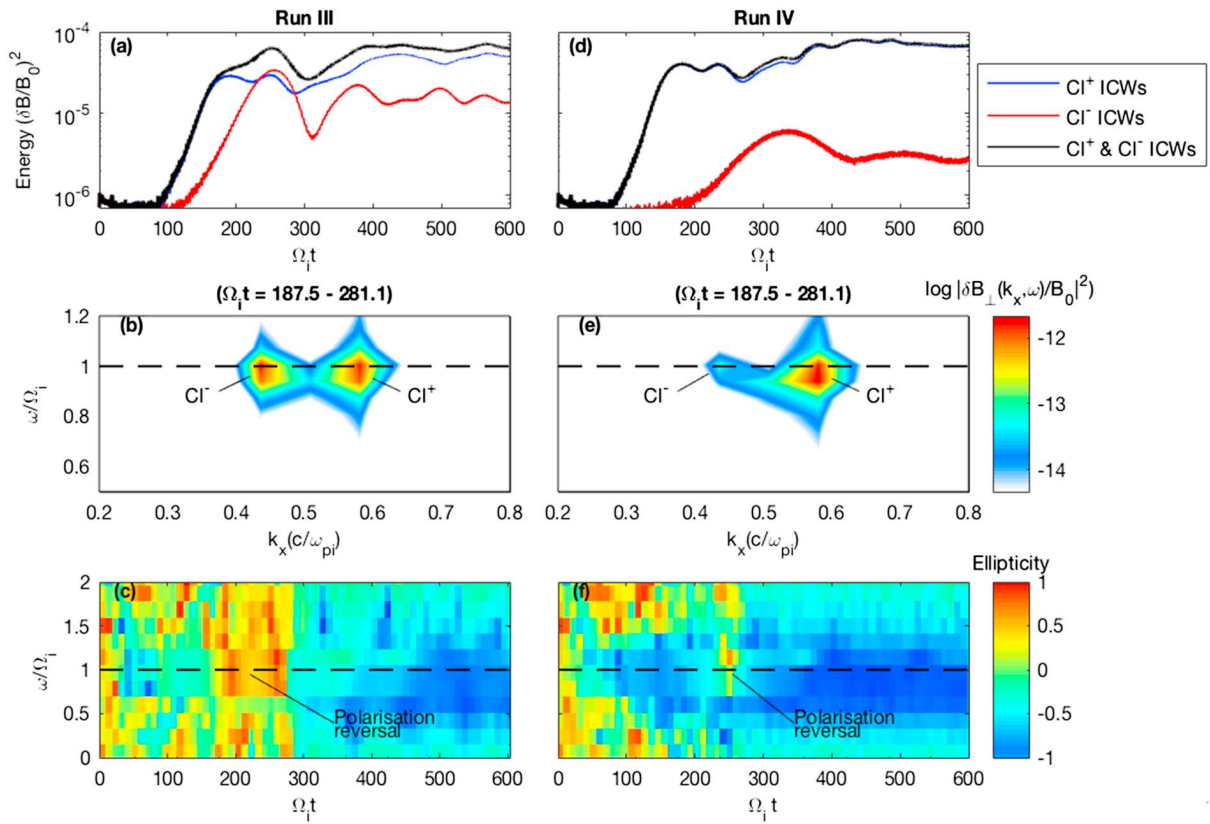

**Figure 9.** Simulation results for Run III and Run IV. (a and d) The fluctuating magnetic field energy densities for the Cl$^+$ (blue), Cl$^-$ (red), and both Cl$^+$ and Cl$^-$ instabilities (black), as outlined in Table 2. (b and c) The frequency-wave number spectrum near the chlorine ion gyrofrequency (dashed black line) for both the Cl$^+$ and Cl$^-$ instabilities, calculated using a Fourier window over $187.5 < \Omega_i t < 281.2$. (e and f) The magnetic polarization for both Cl$^+$ and Cl$^-$ instabilities.

**Table 3**
*Parametric Study of the Scaling of the Chlorine Ion Ring Instability in the Low-Density Regime Applicable to the Galileo E11 and E15 Encounters With Europa*

| Components | Density (1/cm$^3$) | $n_r/n_0$ | Pickup velocity (km/s) |
|---|---|---|---|
| Light ion core | 95 | — | — |
| $Cl^+/Cl^-$ core | 0, 1.9–0.1 | - | - |
| $Cl^+/Cl^-$ ring | 0.1–2, 0.1–1.9 | 0.05, 1 | 55, 100 |

*Note.* The parameter $n_r$ represents the ring ion density, and $n_0$ represents the total chlorine ion density. Further plasma parameters are given in Table 1.

of a varying core population where the chlorine ring and core ion densities sum to 5%, and for zero-core damping. The amplitude modulations brought upon by gyrophase bunching are not included in the saturation amplitudes as a running average was used. This effect would complicate the direct interpretation of the wave amplitudes and introduce a factor of 2 additional uncertainty. The pickup or ring velocity is also varied across the anticipated range observed near the moon (Kivelson et al., 2009), although during E11 and E15 this was in the upper end of this range (Volwerk et al., 2001).

Linear theory predicts the growth rate of the instability is proportional to the ring velocity and increases as the ring-to-core relative density increases (Huddleston et al., 1997). Larger amplitude waves are therefore generated due to the increased amount of free energy available, although linear theory is unable to predict at what amplitudes the instability will saturate. Previous studies have, however, constrained the relationship between saturation amplitudes and ring density and anisotropy (e.g., Cowee et al., 2006; Fu et al., 2016; Huddleston & Johnstone 1992), which is well reproduced in Figure 10 for the negative ion ring instability when relating the varying ring energy to the saturation energy levels, $(\delta B/B_0)^2$.

From this parameter space the $Cl^+$ and $Cl^-$ total maximum densities are constrained to a range of 0.1–1.5/cm$^3$ required to produce the 2.5 and 3.5 nT peak values observed during the E11 and E15 Galileo flybys through Europa's wake. This spread is dependent on the unknown background chlorine ion density and exact pickup velocity.

$Cl^+$ densities have been placed at between 1 and 2.5% of the Io plasma torus (Feldman et al., 2001; Küppers & Schneider, 2000) and highlighted as plausibly present in neutral, negatively, and multiply charged states, as well as produced from Io within compounds such as $FeCl_2$, $CuCl$, $ZnCl_2$, $NaCl$, $KCl$, and $MgCl_2$ (Brown & Hand, 2013; Fegley & Zolotov, 2000). These compounds will diffuse outward toward Europa and likely become implanted within the surface ice before being sputtered into the near-Europa plasma environment and torus, in addition to any local source. $Cl^+$ and $Cl^-$ ions produced locally at Europa would also contribute to the Europan torus and may also contribute to wave damping. The chemical pathways describing the $Cl^+$ and $Cl^-$ production and loss processes are, however, beyond the scope of this analysis, and the results are parametrized with respect to the unknown background chlorine ion densities. It should, however, consequently be noted how important it is to constrain the iogenic material diffusing outward in Jupiter's magnetosphere to understand processes occurring locally at Europa.

The upper limits on the inferred peak $Cl^+$ and $Cl^-$ pickup densities appear high for a sputtered minority species ultimately derived from Io. NaCl has, however, been suggested to be abundant within Europa's subsurface ocean (Kivelson et al., 2000; Zimmer et al., 2000) and could therefore feasibly be present in significant quantities within the Europan ice and plumes. A recent study by Hand and Carlson (2015) also suggest that the characteristic brownish streaks on the moon's surface are composed of high quantities of heavily irradiated NaCl, discolored as a result of the intense >500 krad Europan radiation environment. A Na cloud has also been observed to envelope the moon, and, although Na can be produced through other processes as observed at Mercury (McGrath et al., 1986), significant quantities of Cl ions in localized regions are not inconsistent with physical processes inferred to be occurring at the moon.

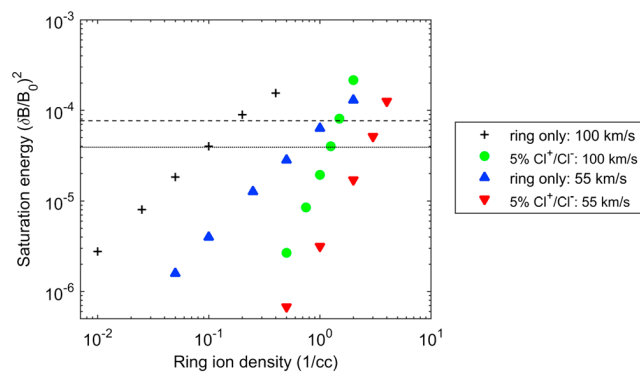

**Figure 10.** Parameter space for the limiting cases of 100 km/s (magenta) and 55 km/s (cyan) pickup velocities. These two cases are simulated with a 5% total ring + core $Cl^{+/-}$ ion density (circles) and with zero-core damping (squares). The $Cl^+$ and $Cl^-$ instabilities have equivalent properties; see section 3. The Galileo maximum wave amplitudes during E11 (dash-dotted line) and E15 (dotted line) are overlaid. The input parameters are provided in Tables 1 and 3.

The LH and RH Galileo wave observations are too brief to directly establish the respective amplitudes for the positive and negative ion instabilities. To interpret these further, Kivelson et al. (2009) show the magnetic field rotations using 15 s hodograms, indicating that the RH wave power is detected over the order of this time period. For the parameter space explored, the dynamic spectra showed evidence of both LH and RH polarizations for density ratios as low as 1:19 $Cl^-$:$Cl^+$. Below this lower limit of ∼5% the $Cl^-$ instability was not sufficiently strong to produce a polarization reversal.

## 6. Summary and Conclusions

This study used a hybrid simulation technique to characterize the physics of ICWs generated by anisotropic distributions of positive and negative pickup ions. These simulations have verified the predictions of linear dispersion theory and were used to relate wave amplitudes and wave polarizations observed by Galileo to local ion populations. Through the characterization of the negative ion ring driven instability and the simultaneous generation of the positive and negative ring instabilities within the same system, the following were concluded:

1. A negatively charged pickup ion ring will generate near circularly polarized RH waves consistent with theoretical predictions.
2. The RH ring instability behaves analogously to the LH ring instability with respect to the resonant frequencies, wave numbers excited, and interactions with a gyroresonant core population.
3. The presence of both RH and LH instabilities act to bunch both ring and core ion populations which results in an additional and potentially significant factor of ≤2 modulation to the wave amplitudes.
4. The presence of both positive and negative ring ions results in variability in ICW polarization with the dynamic spectra displaying LH, RH, and linear polarization, the latter as anticipated for oppositely handed circularly polarized waves with otherwise equivalent spectral properties.
5. Despite the modulations associated with gyrophase bunching, the positive and negative ring instabilities behaved independently as indicated by the time histories of their average wave amplitudes, component temperatures, and wave spectral properties.

After characterizing the behavior of the instability in the Europan plasma environment, the simulations and Galileo wave observations were compared to determine if $Cl^-$ pickup ions are able to explain the wave signatures observed by Galileo and if it is possible to estimate the absolute densities of the chlorine ions from the observed waves. The following were concluded:

1. The simulations and linear theory demonstrate that for the Europan pickup geometry, the RH wave power observed by Galileo is consistent with the pickup of $Cl^-$.
2. The simulations suggest that variability in the wave polarizations is observed in systems which contain instabilities driven by positive and negative ion rings, as was observed by Galileo.
3. From the overall E11 and E15 wave amplitudes the $Cl^+$ or $Cl^-$ densities are estimated to peak in the range of 0.1–1.5/cm$^3$, this spread dependent on the unknown background $Cl^+$ and $Cl^-$ densities and exact pickup velocity.
4. The ICWs at the chlorine ion gyrofrequency observed by Galileo were too brief to establish amplitudes for the LH and RH waves individually, and a threshold of ∼5% of the total chlorine pickup ions was established, below which an observable signal of the $Cl^-$ instability would not be present.
5. The occurrence of bunched ion populations, caused by the presence of both LH and RH ICWs can, however, complicate interpretation of ICWs driven by positive and negative pickup ions.

This study has investigated the behavior of the RH Alfvén-cyclotron instability and also a method of using the wave polarization to identify the presence of negatively charged pickup ions. This method is envisaged to be relevant to other environments in the outer solar system where negative ions can exist in abundance and also provides context for planned in situ observations to be gained from the upcoming ESA JUpiter ICy moons Explorer (JUICE) mission and the NASA Europa Clipper mission.

**Acknowledgments**

R. T. D. acknowledges a LANL Vela Fellowship, STFC Studentship 1429777 and an RAS grant. R. T. D. and M. C. acknowledge the Los Alamos Space Weather Summer School funded by the Center for Space and Earth Sciences (CSES) at LANL. X. F. was supported by Energetic Particle, Composition, and Thermal Plasma (RBSP-ECT) investigation funded under NASA's Prime contract NAS5-01072. The research effort of S. P. G. was supported by the National Aeronautics and Space Administration grant NNX16AM98G. A. J. C. acknowledge support from the 241 STFC consolidated grant to UCL-MSSL ST/N000722/1. The simulation data in this paper are available by contacting the corresponding author.